\documentclass[%
 reprint,superscriptaddress,
 amsmath,amssymb,nofootinbib,
 aps,twocolumn,prl
]{revtex4-2}
\usepackage[utf8]{inputenc}
\usepackage{bbold}

\newcommand{\eqn}[1]{\begin{eqnarray} #1 \end{eqnarray}}
\newcommand{\tit}[1]{\textit{#1}}
\newcommand{\tbf}[1]{\textbf{#1}}

\newcommand{\ket}[1]{| #1 \rangle}

\newcounter{theorem}
\newenvironment{theorem}[1][]{\refstepcounter{theorem}\par\medskip
   \noindent\textbf{Theorem~\thetheorem. #1} \rmfamily}{\medskip}

\newcounter{proposition}
\newenvironment{proposition}[1][]{\refstepcounter{proposition}\par\medskip
   \noindent\textit{Proposition~\theproposition #1:} \rmfamily}{\medskip}

\newcounter{definition}
\newenvironment{definition}[1][]{\refstepcounter{definition}\par\medskip
   \noindent\textbf{Definition~\thedefinition. #1} \rmfamily}{\medskip}

\begin{document}

\title{A single space-time is too small for all of Wigner's friends}

\author{Jacques L. Pienaar\smallskip}
\affiliation{
QBism Group, University of Massachusetts Boston, Boston MA 02125, USA}
\affiliation{
Instituto de Física, Universidade Federal do Rio de Janeiro, Rio de Janeiro, RJ 21941-972, Brazil}

\begin{abstract}
  Recent no-go theorems on interpretations of quantum theory featuring an assumption of `Absoluteness of Observed Events' (AOE) are shown to have an unexpectedly strong corollary: one cannot reject AOE and at the same time assume that the `observed events' in question can all be embedded within a single background space-time common to all observers. Consequently, interpretations that reject AOE appear incompatible with a `block universe' view of space-time.
\end{abstract}

\maketitle

\section{Introduction}
Recently, a number of `extended Wigner's friend' (EWF) no-go theorems have been proposed~\cite{brukner_2018,frauchiger_quantum_2018,bong_strong_2020,PuseyMasanes_2017,Leifer_2020,EWF_healey_quantum_2018,EWF_leegwater_when_2022,EWF_ormrod_no-go_2022,EWF_allard_guerin_no-go_2021,EWF_wiseman_thoughtful_2023,EWF_haddara_possibilistic_2023,EWF_utreras-alarcon_allowing_2023,schmid_review_2023}. These thought experiments combine elements of the classic ``Wigner's friend" thought experiment\cite{WF_original} with the scenario famously studied by Bell~\cite{Bell_original}. Collectively the new no-go theorems pose a serious challenge to an assumption called the \tit{Absoluteness of Observed Events (AOE)}, which states that the outcome of any measurement performed by an `observer' -- broadly defined here as any entity that can perform measurements -- has a unique definite value that is not relative to the observer who measured it. This has led to increased interest in so-called \tit{perspectival} interpretations, which explicitly reject AOE. Notable examples include QBism~\cite{Fuchs10a,QBism_FDR2014,FuchsStacey2018} and Relational Quantum Mechanics (RQM)~\cite{ROVELLI_96,SEP_RQM}, among others~\cite{Healey_2021,Zwirn_ConSol,dieks_perspectival_2022}. 

In this article we focus on the problem of defining a background space-time in perspectival interpretations. The problem was concisely put by Cavalcanti in a 2021 article on QBism's account of Wigner's friend, in which he wrote:
\begin{quote}
    If we reject AOE [\dots] the classical notion of event must also be challenged. In this sense the events that are definite for the friend but not for Wigner could be said to not be in “Wigner’s space-time”, but occurring in a “Wigner bubble”. --Ref~\cite{cavalcanti_bubble_2021}, p26 (2021).
\end{quote}
If Cavalcanti is right, then on a perspectival interpretation of quantum mechanics, Wigner and his friend cannot regard themselves as embedded within a single space-time that encompasses all events that occur relative to both of them.
Among perspectival interpretations, only QBism has explicitly embraced the idea that different observers do not inhabit one and the same space-time. For evidence of this, one need look no further than the title of Fuchs' 2014 samizdat (``My Struggles with the Block Universe"~\cite{fuchs_samizdat_2015}); however the same sentiment can be found in Fuchs' correspondences as early as 2007~\cite{cavalcanti_bubble_2021,fuchs_PC_2007}. Specifically, QBism regards space-time primarily as an ``abstract diagram" that any agent can use as a guide to co-ordinating their own expectations with their own spatio-temporal movements~\cite{mermin_commentary_2014,QBism_FDR2014}; in particular it should \tit{not} be understood as an objective description of reality independently of the agent's activity. Thus, the fact that multiple agents use the same diagram only implies (at most) that these agents' experiences individually relate to space-time in a similar way; it does not imply that they literally inhabit one and the same space-time `arena' in which all space-time events may be treated as existing relative to all agents. On the contrary, QBism takes the lesson of the Wigner's friend thought experiment to be that such unification is general not possible; as Fuchs wrote in 2011, ``The key lesson [of Wigner's friend] is that each part of the universe has plenty that the rest of the universe can say nothing about. That which surrounds each of us is more truly a pluriverse"~\cite{Fuchs_Schlosshauer}. 

While QBism may have long rejected the idea of a single space-time common to all, it is unclear whether other perspectival interpretations would agree. Moreover, Cavalcanti's argument quoted above is not sufficient to establish that they \tit{must} agree with QBism. 

The reason -- to be explained in more detail in the next section -- is that Cavalcanti's argument only works if the space-time point at which the friend's outcome occurs cannot in principle be identified with any space-time point in Wigner's manifold. This entails a non-trivial assumption about whether and how it is possible to identify space-time points between distinct `possible worlds', and one could potentially escape Cavalcanti's argument by challenging this assumption. 

The main contribution of the present work is to show that even if one is extremely permissive about identifying space-time points between distinct possible worlds, one can always construct an EWF scenario in which Cavalcanti's `Wigner bubble' conclusion still holds. That is: rejecting AOE implies that at least some space-time points cannot belong to a common manifold. Perspectival interpretations therefore appear to be faced with the fragmentation of space-time itself. Whether this is a \tit{reductio ad absurdum} for these interpretations, or represents a new path towards quantum gravity, is now an open question.


\section{Wigner's diamond}

As straightforward as Cavalcanti's argument seems at first glance, on reflection it is not completely obvious that the conclusion follows from the premise. To illustrate how the argument might fail, in this section we introduce a thought experiment called ``Wigner's Diamond", which shows how Cavalcanti's argument is deeply linked to assumptions about the metaphysics of space-time.

First, as a warm-up exercise, let us recall a standard formulation of the ``Wigner's friend" (WF) thought experiment: Wigner's friend measures a qubit inside a sealed laboratory at a pre-arranged time, while Wigner remains outside the laboratory. Assuming that it is valid to apply quantum theory to the joint system of the friend plus the qubit, it is argued that Wigner would be justified in assigning this system a pure entangled state. On the other hand, one could argue that since the measurement actually results in a well-defined unique value for the friend, then it must have a unique value for Wigner too, albeit unknown to him, in which case Wigner should assign a state that is a mixture of the possible outcomes weighted by their respective probabilities. However the latter state will inevitably be a mixed state, so it is inconsistent with the premise that Wigner can assign a pure state to the joint system. We have a contradiction: which is the `correct' way for Wigner to apply quantum theory in this situation?

An important ingredient in deriving this contradiction is the idea that if the outcome exists and has a unique value for the friend, then it must exist and have a unique value for Wigner, the only difference being that Wigner does not know what this value is. This amounts to assuming the Absoluteness of Observed Outcomes (AOE).

Whle there are many ways to potentially resolve the paradox, in what follows we are concerned specifically with the implications of choosing to reject AOE. This means, in particular, that we shall entertain the idea that while the outcome of the friend's qubit measurement exists and has a unique value \tit{for the friend}, it simply does not exist and hence has no value (known or unknown) \tit{for Wigner}\footnote{There is another way to reject AOE, namely, to allow that measurement outcomes can have multiple values in an absolute sense, as would be the case in ``many-worlds" interpretations. However, it is not clear that the EWF thought experiments have any new implications for that class of interpretations beyond what has already been discussed in the literature, so we do not consider them here.}.


Let us now shift focus from the outcome itself to the \tit{place and time} at which it happens for the friend. In principle, this corresponds to a unique space-time point. We now pose a new question: does this same space-time point exist for Wigner? That is: can Wigner point to some location within his own patch of space-time and say meaningfully ``here is the space-time point at which the outcome occurred for the friend", even though it did not occur for Wigner? 

In ``Wigner's diamond", we aim to show that the answer can be `yes'. Inside a hermetically sealed chamber, over which Wigner has significant quantum control, Wigner's friend opens a small box in which resides a diamond. Wigner himself observes the process through a small window into the chamber; however from his vantage point he is unable to see the diamond. Nevertheless, he is able to discern precisely where the diamond is located inside the chamber, as well as the precise moment at which his friend opens the box to examine it.

A the risk of spoiling the mysterious atmosphere, let us add some technical details to the story. The diamond is in fact quite special, as it possesses a Nitrogen-vacancy center (NV-center) that is used to encode a qubit. After opening the small box which contains it, the friend measures it in the computational basis $\{ \ket{1}, \ket{0} \}$ by shining a green (say 637nm wavelength) laser on the location of the NV-center and observing whether it fluoresces or not, corresponding to the outcomes ``$1$" or ``$0$" respectively.

The sealed chamber is moreover fitted with a lattice of rigid measuring rods, each one with a series of clocks embedded along its length, so that an onlooker such as Wigner can easily ascertain the space-time co-ordinates of localized events that occur in the room.

Normally, such access to the otherwise sealed room would enable Wigner to discover what the outcome of the measurement was, thereby spoiling the thought experiment\footnote{`Spoiled' because then the quantum state of the sealed room and its contents no longer be entangled, and the usual WF argument would not go through.}. This is avoided by the presence of a very small screen -- no larger than the diamond itself -- situated between the diamond and the window, effectively blocking Wigner's view of the NV-center. 

By this and perhaps other means that we shall not trouble ourselves to imagine, we shall assume it can be arranged that no information about the NV-center's fluorescing (or not) is available in the physical fields that escape the laboratory through the window glass; yet at the same time that it is possible for Wigner to clearly see the location of the tiny screen and thereby infer the location of the NV-center qubit within reasonable accuracy, as well as the precise time at which the green laser is shone onto it, which he can infer by looking at a nearby clock\footnote{Our setup is inspired by David Deutsch's famous twist on the WF thought experiment, in which he argues that it is possible for information inside the lab to reach Wigner on the outside without disrupting the coherent entangled superposition of friend-plus-qubit; the caveat is just that the information conveyed must not contain any clues as to \tit{which} outcome was obtained; see Ref.~\cite{DeutschWF}. }.


The suggestion is now clear: the abstract co-ordinate frame marked out by the physical rulers and clocks can be shared between Wigner and his friend, moreover both can agree without a doubt as to where and when the green laser impinged upon the diamond. The only point of disagreement -- which follows from our having rejected AOE -- is whether the qubit measurement actually resulted in an outcome having a definite value. For the friend, the NV-center either fluoresces or it does not, and he is well aware which is the case. For Wigner, the light field at the site of the NV-center has become entangled with the friend, and there is no outcome to speak of. Wigner cannot say that the NV-center has fluoresced, nor can he say that it has not fluoresced, but more profoundly he cannot even say that it must be one or the other: \tit{the value of the friend's outcome simply does not exist for Wigner as an event}. 

Strange though it may be, Wigner has no difficulty at all in answering questions like ``where and when did the outcome occur for the friend?" using the information which is plainly visible and hence shared by both parties. In this situation, it seems difficult to maintain Cavalcanti's conclusion from the Introduction, namely, that the friend's outcome is not localizable anywhere in Wigner's space-time. Evidently it has a place in the friend's space-time at a well-defined point; moreover \tit{the same point} apparently exists also in Wigner's space-time, marked out by the same co-ordinates. Is this a counter-example to Cavalcanti's argument?

\section{Wigner's friend and the metaphysics of space-time}

Perhaps contrary to intuition, the ability for Wigner and his friend to agree on the co-ordinates at which the green laser strikes the diamond is not by itself sufficient to establish that these co-ordinates mark one and the same \tit{space-time point}. Moreover, on at least one widely accepted account, such an identification \tit{cannot} be established -- not even in principle! 
 
To see why, we need to dip into the literature on the metaphysics of space-time, particularly Einstein's hole argument~\cite{SEP_hole} and its more recent extension, the \tit{quantum hole argument}~\cite{QHoleArgument}. While a proper review of these issues is beyond the scope of this article, we provide a brief sketch of the main issues relevant to our discussion.

Within general relativity (GR) it is important to distinguish \tit{space-time points}, namely the abstract mathematical points that comprise the Riemannian space-time manifold, from what we shall call \tit{localized material co-incidences} or for brevity \tit{material events}, which refer to the particular arrangements of matter and metric fields that obtain in the vicinity of a given space-time point. Now, this distinction poses a metaphysical problem, which was first articulated by Einstein in his famous ``hole argument": given one way of distributing the matter and metric fields over the manifold of space-time points, one can perform an active general co-ordinate transformation (diffeomorphism) to obtain a new, equally valid distribution that is observationally equivalent to the first, due to the general covariance of the equations of GR. Consequently, GR provides no way in principle to determine which space-time point `really' corresponds to a given material event. If we wish to regard the two matter-field distributions as representing different states of affairs in reality, then we must admit that solving the equations of GR is insufficient to determine the real state of affairs: this is the problem of `under-determination'.

One response, favoured by Einstein, is to invoke Leibniz's principle of the identity of indiscernibles and demand that where no observational difference obtains, there should be no \tit{metaphysical} difference either. This amounts to saying we should identify the set of space-time points with the full set of material events, for `space-time' as such has no existence beyond the particular relations of metric and matter fields among themselves. 

This view, called the \tit{relational} view of space-time, avoids under-determination but implies that there is no meaningful way to compare alternative configurations of material events at the `same' space-time point. If we adopt the relational point of view in our consideration of Wigner's diamond, the friend's measurement outcome does not merely happen `at' a space-time point, as though inside a container; rather, it contributes to the very \tit{definition} of that point (together with the other fields coincident with it). Moreover, since the outcome in question \tit{does not exist} as a material event for Wigner, it follows that the space-time point that it defines also cannot exist for Wigner. Thus: it is not in Wigner's space-time manifold, recovering Cavalcanti's original conclusion.

What happens if we reject the relational view of space-time? An alternative favoured by some philosophers is \tit{space-time substantivalism}, which holds that space-time points exist quite independently of the matter and metric fields. This approach allows us to compare different configurations of the matter and metric fields at the same space-time point, at the cost of embracing metaphysical under-determination. On this view, nothing prevents us from asserting that the space-time point at which the friend's outcome occurs also exists as a point in Wigner's space-time. 

Recently, the authors of Ref.~\cite{QHoleArgument} have proposed a third option that we call the \tit{reference frame} view of space-time, which lies somewhere between the two extremes. According to this view, space-time points are identified with the material events of a special subset of matter fields, which are called the \tit{reference frame} (RF). Once fixed, the RF provides an operational means of identifying the `same' space-time point between different configurations of the \tit{other} fields not included in the RF. Furthermore it avoids the radical under-determination of space-time substantivalism because the identity of space-time points is still ultimately grounded in matter and metric fields\footnote{It does suffer from a different kind of under-determination due to the inherent ambiguity in the choice of fields that belong to the RF; this is the main conclusion of the \tit{quantum hole argument}~\cite{QHoleArgument}}. 

As with the substantival view, the RF view of space-time allows us to avoid Cavalcanti's argument, \tit{provided that} we choose a suitable RF. For instance, the system of rigid rods and clocks that form the spatio-temporal `grid' in Wigner's Diamond enable both Wigner and the friend to agree \tit{operationally} that the green laser impinges on the diamond at one and the same juncture within the grid. Taking this grid as the RF would then be sufficient to identify this juncture as representing \tit{the same space-time point} for both parties, even though the measurement outcome which happens there for the friend (the fluorescence or non-fluorescence of the NV-center) does not happen at all relative to Wigner. This enables us to make sense of the otherwise peculiar phrase: ``that is the unique point in space-time \tit{shared by Wigner and his friend} where the outcome occurred for the friend \tit{but not for Wigner}".

To conclude, we have shown that if one does not adopt a relational view of space-time, then Cavalcanti's argument is not decisive in forcing us to conclude that there must be a set of space-time points that exist in the friend's manifold but not in Wigner's manifold, \tit{ie} a ``Wigner bubble". On the other hand, the door remains open to a slightly strengthened version of the argument that would be decisive for reaching this conclusion, regardless of one's interpretation of the metaphyics of space-time. We present such a strengthened argument in the next section.

\section{Concepts and definitions}

In the remainder of this article we provide formal definitions that will allow us to prove that the conclusion of Cavalcanti's argument is completely general. That is, regardless of one's interpretation of space-time points, one can always formulate a version of the WF (and EWF) thought experiments such that rejecting AOE is sufficient to prove that there is a ``Wigner bubble": a sector of space-time points that belong to the friend's space-time manifold but which cannot belong to Wigner's space-time manifold. Clearly, we cannot therefore begin by assuming that there is a \tit{single} manifold of space-time points shared between Wigner and his friend; we shall therefore need to carefully define what we mean by space-time and locality before we can proceed to the main result.

The typical scenario in most EWF scenarios involves an \tit{experiment} in which \tit{observers} labeled $A,B,C,\dots$ perform measurements and obtain outcomes over a number of experimental \tit{runs}. During any given run, we assume that each observer chooses to perform just one measurement selected from a set of precisely $N$ possible measurements, and that each measurement has $M$ possible outcomes. We use the following notation for the measurements and their outcomes: the measurement settings of observers $A,B,C,\dots$ are denoted by respective variables $x,y,z,\dots \in \{1,2,\dots N\}$; furthermore the outcome obtained by each of the observers $A,B,C,\dots$ are represented by the respective variables $a,b,c,\dots \in \{1,2,\dots M\}$.


In any given experimental run, we would like to assert that the measurement settings of all observers simultaneously take definite values, that is, we can imagine a list that tells us precisely what values of $x,y,z,\dots$ are chosen in a given run. However, phrased in this way, it is not clear that any actual observer could verify that this assertion is true: it seems to appeal to a `view from nowhere', which is counter to the spirit of perspectival interpretations. To avoid this problem, we shall explicitly include an ultimate observer whom we shall call `the spectator', whose role is merely to collect and record data about all of the other observer's measurement settings. Specifically, we assume that in each run all observers are required to communicate (say, via classical channels) their choices of setting variables $x,y,z,\dots$ to the spectator, who then maintains a record of their values, over all runs, until the end of the experiment.

A very significant feature of the EWF no-go theorems is that they are constructed in such a way that the observers' \tit{outcomes} cannot all be communicated to the spectator in this manner. This is because the EWF thought experiments, following the same general pattern as some variants of the original Wigner's friend thought experiment, involve witnessing ``coherent interference" of some observers' measurement outcomes by measurements performed on them by other ``super-observers". This process -- however one chooses to interpret it -- has the practical consequence that the ``which-outcome" information of certain outcomes is not available to our external spectator during the course of the experiment. If this information \tit{were} to be conveyed to the spectator, it would \tit{spoil} the experiment in the sense that the relevant no-go theorem would no longer apply. The experimental protocol therefore requires that the observers must each keep their measured outcomes to themselves and not communicate them to anyone else. 

Consequently, in the kind of experiments that we are considering, there is no single observer (not even the spectator) who has access to a complete record of all the measurement outcomes of all observers in any given run of the experiment.

Despite this limitation, if one is not committed to adopting a perspectival interpretation, then one has the option of positing a `view from nowhere' in which all the outcome variables $a,b,c,\dots$ of all observers \tit{do} simultaneously have definite unique values in each run (never mind that no actual observer can know what they are); this is precisely the assumption of \tit{Absoluteness of Observed Events} (AOE).

The assumption AOE is formulated explicitly in reference to measurements on quantum systems and their associated outcomes relative to the observers who perform them. As such, accepting or rejecting AOE does not have any \tit{direct} implications for how one regards space-time itself. In particular, one might find it very natural to assume that while the observers in the EWF experiments do not necessarily share the outcomes of their measurements, they do nevertheless share the same space-time as one another. Let us try to make this the idea of ``sharing the same space-time" more precise.

As previously noted, a \tit{space-time point} is a point in a smooth and differentiable (hence `classical') manifold $\mathcal{M}$, which can be parameterized in any local patch by co-ordinates of the form $x^{\mu} := (x^0,x^1,x^2,x^3)$, where $x^0$ conventionally represents the time co-ordinate. In order to make room for alternative interpretations of space-time points (per the discussion in the preceding section) we do not assume space-time relationalism, \tit{ie} we do not assume that the identity of a space-time point is exhausted by the configuration of all matter and metric fields that coincide in its vicinity.


When an observer (say $A$) observes an outcome, thereby establishing its definite unique value relative to \tit{at least} $A$ himself, we shall assume that this occurs in the neighbourhood of some specific space-time point $e \in \mathcal{M}_A$, where $\mathcal{M}_A$ is a manifold that represents a local region of space-time that is associated to $A$. (The precise definition of $\mathcal{M}_A$ is flexible: it can be fulfilled by any manifold that is sufficiently expansive to include the locations of all events which are deemed to occur \tit{for} $A$). The assumption that each outcome occur at a \tit{specific} fixed space-time point will be called \tit{relative locality} of outcomes for $A$; it can be achieved by locally `fixing a gauge', that is, picking out a preferred way of distributing the matter and metric fields which are accessible to $A$ over the manifold $\mathcal{M}_A$.

Let us now restrict attention to (say) $A$'s outcome variable $a$, and let $\mathcal{O}_a$ be the set of all values that $a$ could possibly take in any given experimental run. According to the assumption of relative locality, there must be a map $\mathcal{L}_A:\mathcal{O}_a \mapsto \mathcal{M}_A$ which assigns each value in $\mathcal{O}_a$ to the corresponding space-time point at which it occurs for $A$, in any run where it does occur for $A$. We shall refer to this as $A$'s \tit{localization map}. Let $\epsilon_a \subset \mathcal{M}_A$ denote the image of the set $\mathcal{O}_a$ under the map $\mathcal{L}_A$. The same definitions as we have just made for observer $A$ are to be made for all other observers $B,C,\dots$, \tit{mutatis mutandis}.

Building on the foregoing definitions, we can now formalize what we mean by saying that an experiment can be embedded in a `block space-time', or more precisely, that all outcomes can be embedded within a single, shared space-time manifold: \\

\tbf{Definition:} We say that the \tit{classical background space-time condition (CB)} holds for an experiment iff there exists an embedding of all observers' manifolds $\mathcal{M}_A, \mathcal{M}_B, \mathcal{M}_C, \dots$ as submanifolds of some single manifold\,\footnote{Of course, in a perspectival interpretation, $\mathcal{M}$ is implicitly itself the space-time of an observer, namely the spectator; but if CB holds then we can drop the indexical and take $\mathcal{M}$ to be simply `the' manifold for all observers.} $\mathcal{M}$.\\

Now, in order to evade the conclusion of Cavalcanti's argument, we must be able to show that all the outcomes in an EWF thought experiment can be modeled as occurring within the same space-time manifold $\mathcal{M}$. At minimum, this requires that there exists a map $\phi$ which associates the outcomes obtained in each run to their respective space-time locations in $\mathcal{M}$. We formalize this requirement by the following definition:\\

\tbf{Definition:} Let $\mathcal{O}_{abc\dots} := \mathcal{O}_a \cup \mathcal{O}_b \cup \mathcal{O}_c \dots$ denote the union of the sets of possible outcome values for all observers in any run of the experiment. Consider a map $\phi: \mathcal{O}_{abc\dots} \mapsto \mathcal{M}$ that takes each set of possible outcome values for all observers to corresponding set of space-time points in $\mathcal{M}$. We say that $\phi$ is \tit{consistent} with a given localization map, say $\mathcal{L}_A$, if it assigns distinct space-time points to the outcome values of $a$ whenever $\mathcal{L}_A$ does. That is, $\phi$ is \tit{consistent} with $\mathcal{L}_A$ iff $\forall a',a'' \in \mathcal{O}_a$: 
\eqn{ \label{eqn:phi_consistency}
 \mathcal{L}_A(a') \neq  \mathcal{L}_A(a'')\Longleftrightarrow  \phi(a') \neq \phi(a'') \, . 
}

If $\phi$ is consistent with the localization maps of all observers in the experiment, we call it an \tit{absolute localization map} for that experiment.\\

The preceding definition can be motivated as follows: intuitively, the existence of an absolute localization map $\phi$ means that in any given run of the experiment, it is \tit{possible} to assume that every observed outcome happened at some definite space-time location, as given by the map. Moreover, consistency with the individual observer's perspectives demands that whenever (say) observer $A$ finds that different values $a',a''$ occur at different space-time points in his own manifold patch $\mathcal{M}_A$, the map $\phi$ should preserve this fact by mapping $a',a''$ to distinct space-time points in the global manifold $\mathcal{M}$.

\section{Main result}

With the basic definitions of the preceding section now in place, we can proceed to prove the main result. We begin with the following assumption, that we will argue is quite reasonable:\\

\tbf{Assumption:}  It is possible for any observer to implement their quantum measurement in such a way that different outcome values occur at different space-time points relative to themselves. Consequently, their localization map is injective (\tit{ie} ``one-to-one"). \\

This might seem so obvious as to not merit being labeled an assumption. For example, suppose the qubit to be measured is encoded in the polarization of a single photon. Detection then involves passing it through a polarizing beamsplitter to one of two detectors situated in different locations, and the location of the detector that clicks indicates the value of the measured polarization. This would seem to guarantee that different measurement outcomes occur at different space-time points, at least relative to observer performing the measurement. 

However, in light of the discussion about the metaphysics of space-time points, we must not leap to conclusions. In fact, since the two outcomes represent mutually exclusive alternative events (either one clicks, or the other one does), asserting that these events occur \tit{at different space-time points} is just as problematic as asserting that they occur at the \tit{same} space-time point, because either way the assertion requires some method of identifying space-time points between different possible worlds. Indeed, if we subscribe to a relational view of space-time, the assertion is meaningless. Thus, in making this assumption, we are already implicitly adopting either the RF interpretation or a substantival interpretation of space-time. Fortunately it is precisely in these cases where Cavalcanti's conclusion might be challenged, so we have nothing to lose by making this restriction.

To properly justify the assumption, then, let us return to the example of the polarization qubit. Suppose that, contrary to our assumption, the two material events in question (detector A clicking, or detector B clicking) just so happen to be associated with one and the same space-time point. If true, this fact would have to be established either by appealing to a material RF (in the reference-frame interpretation) or by fixing a gauge that specifies which material events correspond to which space-time points (in a substantival interpretation). In either case, we can always make our assumption true by adjusting the experiment slightly so as to ensure that the two material events occur at \tit{different} space-time points -- for instance by moving one of the detectors to a new location relative to the fixed choice of RF or gauge\footnote{The only way to escape this would be to define the RF or gauge in a pathological way, such as by having all material events occur at a single space-time point. However, such radical maneuvers are evidently self-defeating.}.

Proceeding to our proof, the first step is to show that the existence of an absolute localization map $\phi$ is a necessary condition for CB to hold. More precisely:\\

\tbf{Proposition:} If CB holds for an experiment, then we can construct an absolute localization map for the experiment.\\

\tit{Proof:}
Since all observers' manifolds can be embedded in a single background space-time $\mathcal{M}$ (per CB), there is a `canonical' way to define $\phi: \mathcal{O}_{abc\dots} \mapsto \mathcal{M}$, namely as the map which takes $a \mapsto \mathcal{L}_A(a)$, $b \mapsto \mathcal{L}_B(b)$, \dots \tit{etc}, for all values $a \in \mathcal{O}_a$, $b \in \mathcal{O}_b$, \dots; in other words, $\phi$ is just the set of localization maps, composed in parallel. Such a map is well-defined since CB guarantees that its output domain will be a subset of $\mathcal{M}$. Specifically, the output domain of $\phi$ is the union of the sets of points in $\mathcal{M}$ that are the outputs of the relevant localization maps: $\mathcal{E}_{abc\dots} := \epsilon_a \cup \epsilon_b \cup \epsilon_c \cup \dots$, and $\mathcal{E}_{abc\dots} \subset \mathcal{M}$. Thus by construction $\phi$ is a valid absolute localization map. $\Box$ \\

\tbf{Main result:} Consider an EWF scenario in which accepting some set of assumptions -- it doesn't matter what they are specifically -- would force one to reject AOE. Then one can design a similar version of this scenario in which the same set of assumptions would imply that there cannot be an absolute localization map (hence that CB must be rejected too).\\

\tit{Proof:} It will be useful to define a \tit{cross-section} of $\mathcal{O}_{abc\dots}$ as any subset $\mathcal{X} \subset \mathcal{O}_{abc\dots}$ that contains exactly one value for each of the outcomes $a,b,c,\dots$. Call a cross-section \tit{valid} if it represents a possible assignment of values to the outcomes $a,b,c,\dots$ in any given run of the experiment under consideration. Note that if AOE holds, there must exist at least one valid cross-section.

Next, consider designing the same EWF experiment such that each observer's localization map takes each of their possible outcomes to a distinct space-time point in the observer's local manifold. By the assumption at the beginning of this section, this is always possible. If an absolute localization map $\phi$ exists for this experiment, it must assign a space-time point to each of the outcomes $a,b,c,\dots$, in each run. However, consistency (Eq~\ref{eqn:phi_consistency}) implies that $\phi$ is a bijection which maps between the set of points $S$ assigned in a given run, and the values of the outcomes obtained in that run. It follows that $\phi^{-1}(S) \subset \mathcal{O}_{abc\dots}$ is a valid cross-section. 

The result is now immediate: for if we adhere to any set of assumptions that jointly prohibit AOE, these same assumptions also prohibit the existence of any valid cross-section; therefore there can be no absolute localization map for this experiment.  $\Box$ \\ 

Since (as we showed earlier) an absolute localization map is a necessary precondition for the `block space-time' condition CB, any EWF scenario can be modified in the manner described above to ensure that the very set of assumptions which contradict AOE will also contradict CB. To the extent that EWF experiments challenge AOE, therefore, they also challenge the supposed unity of the space-time manifold itself; moreover this is true regardless of one's commitments as to how space-time points should be identified in principle.

\section{Discussion and Conclusions} 

We remark on three main consequences of the foregoing result concerning locality and space-time in perspectival interpretations.

Firstly, many of the EWF no-go theorems include an assumption of ``\tit{locality}", which one might hope to retain by rejecting AOE; after all, one usually only accepts one horn of a multi-horned dilemma on the condition of being able to avoid the other horns. However, our result shows that for no-go theorems where `locality' is defined using terms that presume CB, such as `space-like separation' and `light cones', one cannot retain the relevant notion of locality by rejecting AOE.

Although perspectival interpretations cannot therefore be `local' by traditional definitions of that term, it remains possible that they may be compatible with some more nuanced conception of locality that does not require CB. In particular, while both QBism and RQM have claimed to be `local' in some sense~\cite{QBism_FDR2014,martin-dussaud_notion_2019}, these claims have been disputed by others~\cite{cavalcanti_bubble_2021,pienaar_comment_2019}, with the criticisms leveraging the fact that the definition of `locality' becomes ambiguous without AOE. 

The most direct way to answer such criticisms is therefore to provide an explicit definition of `locality' that does not require AOE. One strategy which appears natural in light of the present work would be to define ‘locality’ as the requirement that \tit{relative locality} holds for each observer (as defined in the previous section). Accordingly, space-time concepts should always be indexed to a specific observer, for instance we may take the ``space-time of observer $A$" to refer to the manifold $\mathcal{M}_A$ equipped with some metric, such that space-time intervals are only defined between
events that occur relative to $A$ and which are therefore embeddable in $\mathcal{M}_A$.

Secondly, the reader might well be alarmed that interpretations of ostensibly \tit{non-relativistic} quantum theory should have led us to radically revise our notion of space-time, especially since the thought experiments leading us to that conclusion all tacitly assume that quantum gravity effects are negligible. However, this may turn out to be an unexpected feature, rather than a flaw, if it turns out that these questions can be answered within the context of a definite proposal for a quantum structure of space-time. In particular, we speculate that there may be a connection to a currently active field of research in quantum gravity based on a principle of `relative locality' (see \tit{eg.}\ \cite{amelino-camelia_principle_2011,amelino-camelia_relative_2011,freidel_modular_2017}); this remains to be explored.

Finally, this result drives home an important distinction between the type of perspectivism that rejecting AOE entails, versus the merely \tit{benign} type of perspectivism that pervades science and is exemplified by special relativity. The latter type posits a model of the world independently of perspectives (\tit{eg} co-ordinate-free Minkowski space-time), and from it derives the contents of any particular perspective by supplying details about how the relevant observer is situated within the world (\tit{eg} a particular inertial reference frame). 

By contrast, both QBism and RQM have emphasized that rejecting AOE places these interpretations beyond the grasp of this benign brand of perspectivism, because it implies that one cannot \tit{in principle} posit a model of the world independently of any perspective. This has led some authors to coin the term \tit{radical perspectivism} in order to emphasize the difference. 

Still, until the present article one might have reserved some doubts as to whether QBism or RQM really necessitate rejecting a block universe. For example, writing about `participatory realist' interpretations like QBism, Dean Rickles~\cite{Rickles_johntology}, citing work by Jenann Ismael~\cite{Ismael_2016}, states: 
\begin{quote}
    [O]ne purported implication [\dots] is that the block universe picture cannot be true, for that gives the world ``once and for all", while here the world is built up from participator interventions. This strikes me as wrong-headed: one can well encompass this core idea within a block universe, just as one can encompass free will in a block by making the switch from an allocentric (block) to an egocentric (embedded) perspective (\tit{cf}.\ Ismael 2016) –
\end{quote}

The EWF no-go theorems, augmented by the arguments we have made here, together decisively close the door on this hope. At the same time, a new door is opened onto the possibility of alternative mathematical structures that can accommodate a more radical form of space-time perspectivism, beyond `the block'.\\

\tit{Acknowledgements:} This work was made possible through the support of Grant 62424 from the John Templeton Foundation. The opinions expressed in this publication are those of the author and do not necessarily reflect the views of the John Templeton Foundation. Thanks to John DeBrota, R\"{u}diger Schack, Marcus Appleby, Blake Stacey, and Chris Fuchs for feedback on an early draft.


%

\end{document}